%% file: sktimevariation.tex
\documentclass[aps,prl,reprint,superscriptaddress,amsmath, amssymb,nobibnotes]{revtex4-1}
\usepackage{graphicx}\usepackage[mathlines]{lineno}\usepackage{amssymb}\usepackage{amsmath}\usepackage{xcolor}\usepackage{relsize}

\begin{document}
\title{Search for Periodic Time Variations of the Solar $^8$B Neutrino Flux\\ between 1996 and 2018 in Super-Kamiokande}
\include{authors-20230602}
\date{\today}
\begin{abstract}
{We report a search for time variations of the solar $^8$B neutrino flux using 5804 live days of Super-Kamiokande data collected between May 31, 1996, and May 30, 2018. Super-Kamiokande measured the precise time of each solar neutrino interaction over 22 calendar years to search for solar neutrino flux modulations with unprecedented precision. Periodic modulations are searched for in a dataset comprising five-day interval solar neutrino flux measurements with a maximum likelihood method. We also applied the Lomb-Scargle method to this dataset to compare it with previous reports. The only significant modulation found is due to the elliptic orbit of the Earth around the Sun. The observed modulation is consistent with astronomical data: we measured an eccentricity of (1.53$\pm$0.35)\%, and a perihelion shift of ($-$1.5$\pm$13.5) days.
}

\end{abstract}
\maketitle

\begin{table*}[ht!]
\centering
\caption{Super-Kamiokande (SK) operation period, live time, energy range, total flux, statistical, and systematic uncertainties of the total flux. The neutrino flux is measured from elastic scattering recoil electrons. For SK-III, 3.99 --4.49 MeV is not used; for SK-IV, 3.49--4.49 MeV is not used in the data.}
\begin{tabular}{c| ccccc}
\hline
SK Phase & & Start date $\sim$ End date & ~~Live days~~&~ Energy range [MeV]~ &~Flux ($\phi_\nu$)+(stat.)+(sys.) [$10^{6}$cm$^{-2}$s$^{-1}$] \\
\hline
\hline
SK-I& & 1996-05-31 $\sim$ 2001-07-15 & 1495.7 & 4.49$\sim$19.5 & 2.35\,$\pm$\,0.02\,$\pm$\,0.08\\
SK-II& & 2002-12-10 $\sim$ 2005-10-06 & 791.9 &  6.49$\sim$19.5 & 2.38\,$\pm$0.05\,$^{+0.16}_{-0.15}$\\
SK-III& & 2006-05-23 $\sim$ 2008-08-17 & 548.5 &  4.49$\sim$19.5 & 2.32\,$\pm$\,0.04\,$\pm$\,0.05\\
SK-IV& & 2008-09-15 $\sim$ 2018-05-30 & 2967.7 &  4.49$\sim$19.5 & 2.31\,$\pm$\,0.014\,$\pm$\,0.040\\
\hline
\end{tabular}
\label{tab_skphase}
\end{table*}
{{\it Introduction.}---}Solar neutrino observations are critical for investigating stars and learning about neutrino physics. The observation of solar neutrinos proves that nuclear fusion powers the Sun. The solar neutrinos carry real-time information about the solar core, while photons take a long time to reach the photosphere. Solar dynamics may cause fluctuations in the solar neutrino flux. The experimental confirmation of neutrino oscillations with solar neutrinos solved the long-standing solar neutrino problem (e.g., the observed average flux of $^8$B neutrinos is as expected\,\cite{sk1_full,sk2_full,sk3_full,sk4_full,PhysRevLett.89.011301}), as well as measured oscillation parameters\,\cite{Bahcall_2001,BAHCALL19981,1993ApJ...408..347T}.

In addition, instantaneous changes in the solar magnetic field could modify the neutrino flux if the neutrino has a magnetic moment. If neutrinos are Dirac particles with a nonvanishing magnetic moment, the magnetic field will rotate the neutrino spin orientation. This process is known as resonant spin flavor precession (RSFP)\,\cite{RSFP_Akh,RSFP_Lim,RSFP_Joa}. The process can occur on the inner tachocline, where a strong magnetic field is formed\,\cite{RSFP_Stur}. Because the spin-flipped right-handed neutrinos are sterile, RSFP would reduce the observed solar neutrino fluxes. 
If neutrinos are Majorana particles with a flavor-changing magnetic moment, the solar magnetic field could convert a $\nu_e$ into a $\bar{\nu}_\mu$ or a $\bar{\nu}_\tau$, reducing the observed solar neutrino due to the suppressed elastic scattering cross sections. Therefore, any time variation in the magnetic fields in the Sun could result in modulations in the observed solar neutrino fluxes. 
The rotation profile of the interior of the Sun is known for $R>0.2R_{\mbox{sun}}$ where $R_{\mbox{sun}}$ is the solar radius. The radiative zone of the Sun in $0.3R_{\mbox{sun}}<R< 0.7R_{\mbox{sun}}$ is assumed to be a solid body rotating at a constant rate\,\cite{SolRot_Cou}. The rotation profile suggests the existence of a magnetic field in the radiative zone. The observations of splitting frequencies of the solar neutrino flux modulation can be used to infer the magnetic field. Observing a periodic modulation in the solar neutrino flux would be a significant breakthrough in understanding neutrinos' magnetic properties and the dynamics of the Sun's inside. We investigate the periodicity of solar neutrino fluxes using the observed Super-Kamiokande (SK) solar neutrino data.\\

\par Super-Kamiokande is a water Cherenkov detector in Kamioka, Japan, with a total mass of 50\,kt\,\cite{FUKUDA2003418}. The solar neutrino data in this analysis were obtained from May 31 1996, to May 30 2018, totaling 5803 days of detector live time. Table\,\ref{tab_skphase} summarizes the periods of data samples, live time, energy range, and overall systematic uncertainties of the solar neutrino flux measurements. While the elastic scattering rate of solar $^8$B (99.8\%) and hep (0.2\%) neutrinos in the SK water is expected to be about 300 per day, most resulting recoil electrons are too low in energy to be observable. Accounting for energy threshold, detector efficiencies, and neutrino oscillations, about 20 solar neutrino interactions per day are observed, while about 45 solar neutrino interactions per day are expected. With this high rate, we search for periodic modulations with periods as short as five days. A comprehensive description of the SK detector can be found in references\,\cite{sk1_full,sk2_full,sk3_full,sk4_full}.\\

\par Solar neutrino events produce recoil electrons through neutrino-electron scattering, preferentially aligned with the Sun's direction. We calculate the angle $\cos\theta_{\mbox{sun}}$ for each elastic scattering event between the reconstructed Cherenkov ring direction and the Sun's direction. The data sample is divided into \(N_{\mbox{bin}}=21\) energy bins: 18 energy bins between 5 and 14 MeV ($\Delta E = 0.5$\,MeV), two energy bins between 14 and 16 MeV ($\Delta E = 1$\,MeV), and one bin between 16 and 20 MeV. We perform a maximum likelihood fit to the $\cos\theta_{\mbox{sun}}$ distribution in each energy bin $i$ to the number of solar neutrino interactions $S$ and the numbers of radioactive background events $B_i$\,\cite{sk4_full}. Each of the $n_i$ events in energy bin $i$ is assigned a signal factor $s_{ik}$ and a background factor $b_{ik}$ depending on $\cos\theta_{\mbox{sun}}$: with the probability density function $p$ for signal ($u$ for background) these factors are $s_{ik}=p(\cos\theta_{k},E_{i})$ and $b_{ik}=u(\cos\theta_{k},E_{i})$.

\begin{equation*}\label{eqn:lik_solfit}
\begin{array}{ll}
{\cal L}=e^{-\left(\sum_i B_i+S\right)}\prod_{i=1}^{N_{\mbox{\tiny bin}}}\prod_{\kappa=1}^{n_i}\left(B_i\cdot b_{i\kappa}+S\frac{\mbox{MC}_i}{\sum_j\mbox{MC}_j}\cdot s_{i\kappa}\right).
\end{array}
\end{equation*}
\noindent The $\cal L$ is maximized by optimizing the signal $S$ and the 21 backgrounds $B_i$. MC$_i$ is the number of events expected in energy bin $i$ using the flux and spectrum of $^8$B and {\it hep} neutrinos assuming no neutrino oscillation. The systematic uncertainties of the neutrino flux measurements consist of energy-correlated and uncorrelated errors. The total uncertainty of the flux is obtained by combining both errors. Table\,\ref{tab_skphase} lists the total systematic uncertainty for each SK period. References\,\cite{sk1_full,sk2_full,sk3_full,sk4_full} describe the systematic uncertainties for the solar neutrino flux measurements. The Sudbury Neutrino Observatory (SNO) measured all active flavors of the solar $^8$B neutrino flux and found ($5.25\pm0.2$)$\times10^{6}$\,cm$^{-2}$s$^{-1}$\,\cite{sno_nc}. This analysis used this value as a reference for solar neutrino flux.

\begin{figure*}[ht!]
\centering
\includegraphics[width=0.9\linewidth]{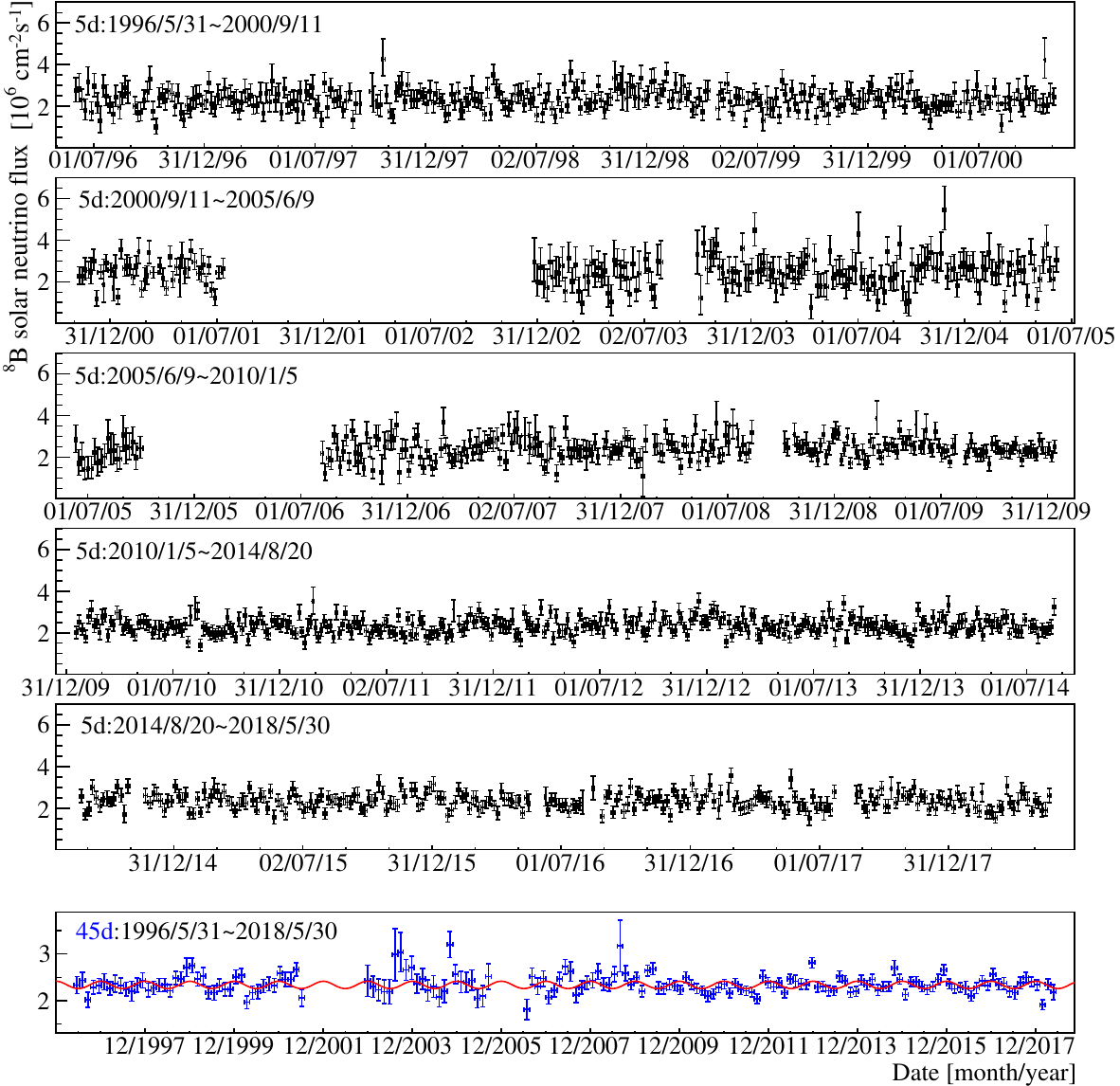}
\caption{Measured $^{8}$B solar neutrino fluxes for five-day (top five panels, black data points) and 45-day (bottom panel, blue data points) intervals without 1/$R^{2}$ correction. The five-day (45-day) plot errors are asymmetric (symmetric) errors of the average fluxes. The solid-red curve in the 45-day plot is the expected sinusoidal solar neutrino flux based on the earth's elliptical orbit.}
\label{solfit_45days_all}
\end{figure*}
\begin{figure}[t!]
\centering
\includegraphics[width=1\linewidth]{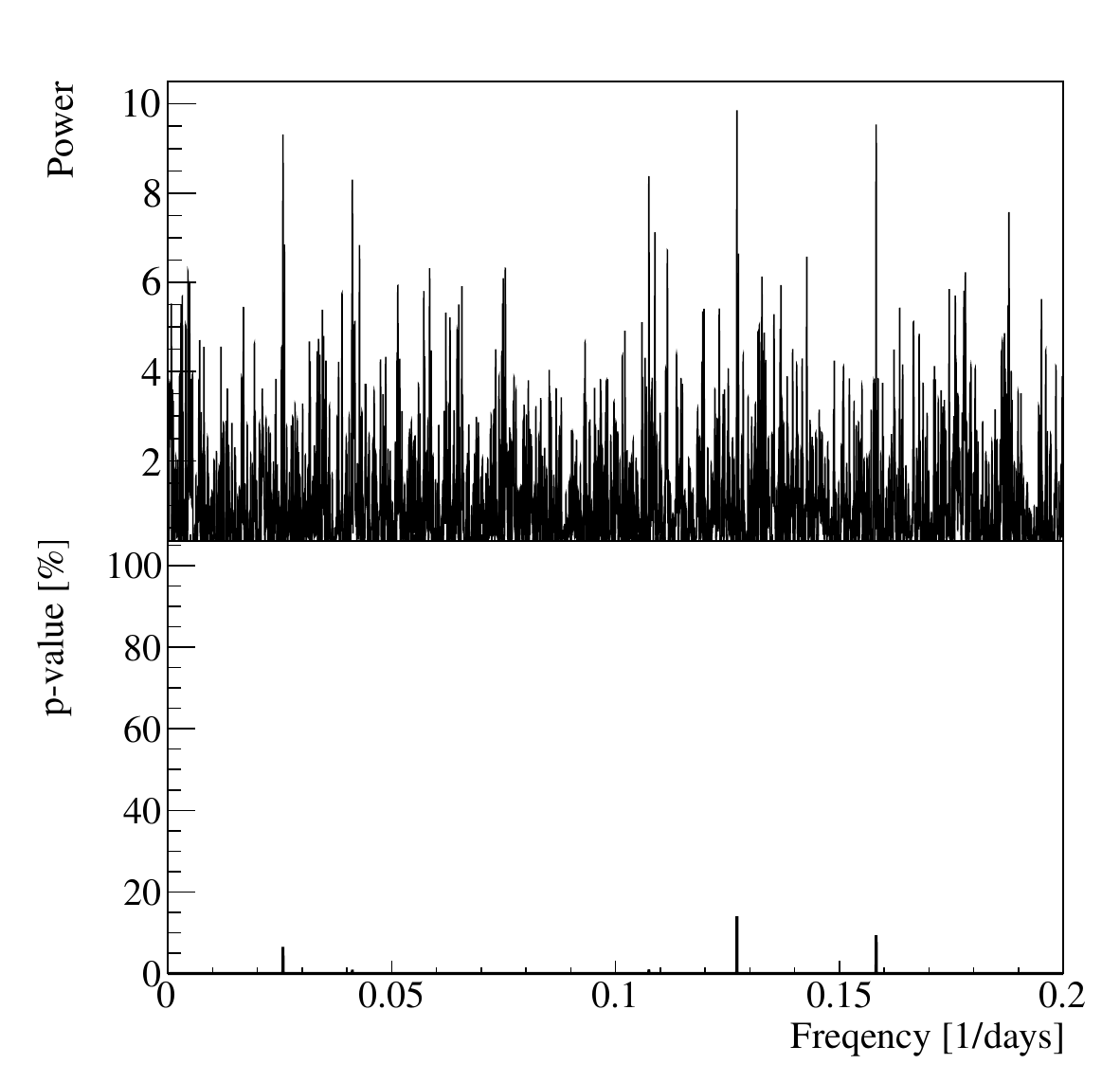}
\caption{Power distribution of likelihood method (top) and $1-p$ (bottom) in percent for $1/R^2$ corrected five-day interval solar neutrino data, which is made by multiplying the original five-day interval data by the squared distance $R^{2}$.}
\label{fig_lik_CL_5d}
\end{figure}

{\it{Search for periodic modulation}.---} To search for periodic variations of the solar neutrino flux, five-day intervals are defined: 358 for SK-I \cite{modu_sk1}, 175 for SK-II, 141 for SK-III, and 669 for SK-IV (total 1343 intervals). The solar neutrino interactions (and uncertainties) are obtained in each interval with the same extended unbinned maximum likelihood method mentioned in the previous section. The flux of $^8$B solar neutrinos is calculated from the observed event rate without assuming the effect of neutrino flavor oscillation. Additionally, no corrections are applied to the flux result for the distance between the Sun and the Earth, which varies annually around 1 A.U. The detector operation time is considered by choosing the weighted mean time as the effective time of the five-day interval. Figure\,\ref{solfit_45days_all} shows five-day and 45-day (average of nine five-day interval data for each bin) intervals of solar neutrino fluxes for all SK phases without corrections for the distance between the Sun and the Earth, which varies annually around 1 A.U. We provide the data for five-day interval fluxes and the measurement time in the Supplemental Material\,\cite{sktimev5804d}. The statistical uncertainty of the five-day interval flux is an asymmetric Gaussian form of the extracted likelihoods. The upper and lower uncertainties of the five-day data point are the displacements from the measured flux at which the profiled likelihood decreases by a factor of $e^{-1/2}$. The data used for the modulation likelihood is corrected by multiplying the squared distance between the Sun and the Earth. The predicted flux in the time bin is obtained by

\begin{equation*}\label{eqn:lik_test}
\begin{array}{ll}
g_{r}(\omega;A,B)=g_{0}+\frac{1}{t_{r,f}-t_{r,i}}\displaystyle\int^{t_{r,f}}_{t_{r,i}}dt\,\left(A\cos\omega t+B\sin\omega t\right),
\end{array}
\end{equation*}
\noindent 
where the frequency ($\omega$) is scanned with a step of $2\times10^{-6}$ days$^{-1}$ and amplitudes ($A$ and $B$) are free parameters, $t_{r,i}$ is the initial and $t_{r,f}$ is the final time of the $r$th time interval. {\it g$_{0}$} is the average of all data points. 
A log-likelihood is defined as, 
\begin{equation*}\label{eqn:chi_lik_test}
\begin{array}{ll}
-\log{\cal L}=\displaystyle \min_{A,B}\left[\frac{1}{2}\sum_{r}\left(\frac{D_{r}-g_{r}(\omega;A,B)}{\sigma_{r}} \right)^{2}\right],
\end{array}
\end{equation*}

\noindent where $D_r$ is the data point at $r$, and $\sigma_r$ is the asymmetric statistical uncertainty $\sigma _{+,r}$ ($\sigma _{-,r}$) for $D_{r}<g_{r}$ ($D_{r}>g_{r}$). The likelihood of the null hypothesis (${\cal L}_0$) is tested by setting the $A=B=0$. 
The power of the likelihood is defined as $P_L=$$-\log({\cal L}_{0}/{\cal L})$. It is assumed that the average of the second term of the $g_r$ equation never deviated from zero for any frequencies. Ten thousand Monte Carlo (MC) samples are generated assuming the null hypothesis to examine a broad range of modulation frequencies. Each MC sample consists of asymmetric Gaussian variations with mean {\it g$_{0}$} and errors given by the upper and lower statistical uncertainties in each interval. Then, the power for each MC set is calculated at all tested frequencies, and the maximum power for each MC set is selected.
	The area of the maximum power distribution less than the corresponding power gives the $(1-p)$ value. Figure\,\ref{fig_lik_CL_5d} shows $P_L$ values and $(1-p)$ for the five-day sample. The maximum power is 9.9 at 0.126 days$^{-1}$ with $(1-p)=$13.7\% which is the area of the distribution below a maximum power of 9.9.
	The results indicate no significant short-term periodic modulation in the measured SK solar neutrino data sample.

\begin{table}[b!]
\centering
\caption{Comparison of time variation study results from SK, SNO, and Borexino experiments. The Kepler constants [eccentricity ($\epsilon$) and perihelion shift ($\delta t_{\text{peri}}$)] and sensitivity limit of amplitude ($A_L$) are from references\,\cite{APPEL2023102778,PhysRevD.72.052010}.} 
\begin{tabular}{c c c c}
\hline
&$\epsilon$ [\%]~~&~~ $\delta t_{\text{peri}}$ [days]& ~~~$A_L$ [\%] (CL)\\
\hline
SK ($^8$B) &~~$1.53\pm0.35$&~~$-1.5\pm13.5$&5.2 (90\%)\\
SNO ($^8$B)&~~$1.43\pm0.86$&-&10.5 (90\%)\\
Borexino ($^7$Be)&~~{$1.84\pm0.32$}&$~~{7\pm20}$&-\\
\hline
\end{tabular}
\label{tab_comp_bosksn}
\end{table}
\par
\begin{figure}[t!]
\centering
\includegraphics[width=1\linewidth]{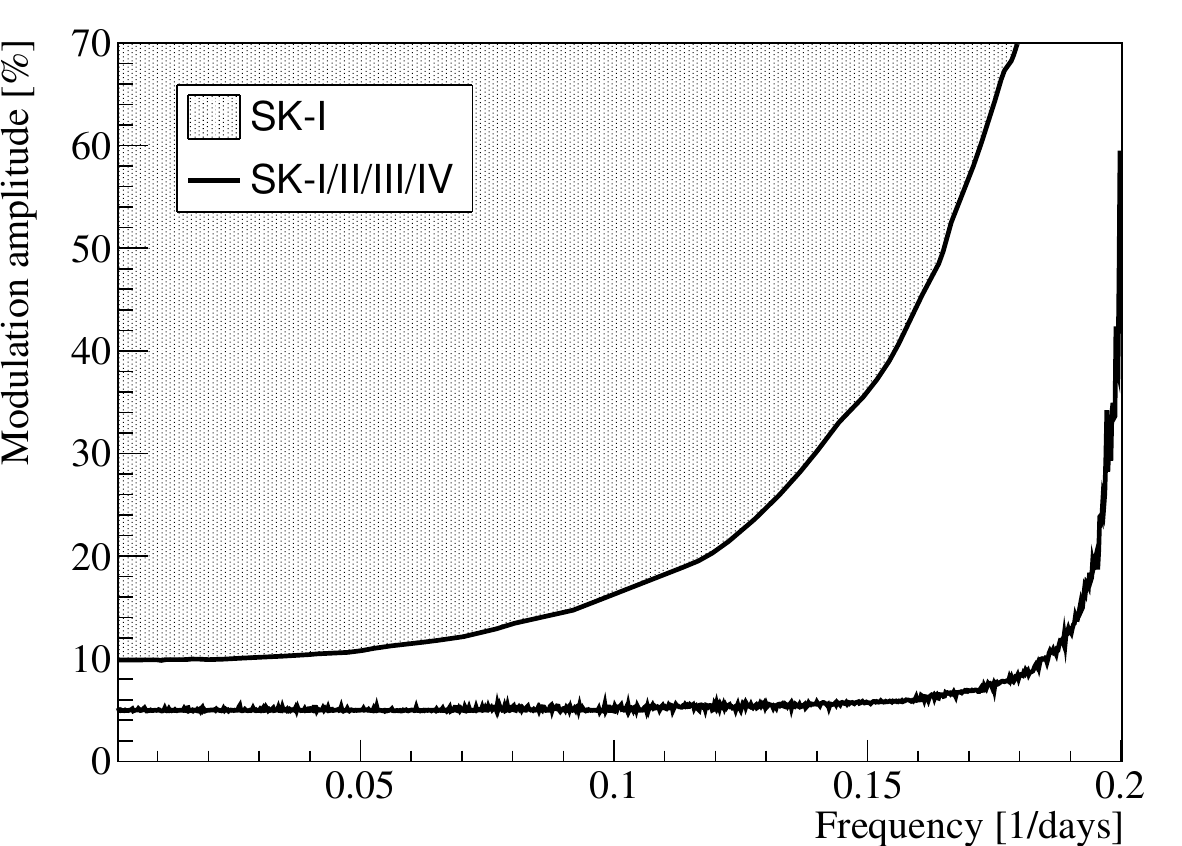}
\caption{The search sensitivity of solar neutrino modulation amplitude for each frequency with 95\,\%CL. The shaded area is ruled out using SK-I solar neutrino data\,\cite{modu_sk1}. The solid curve shows the results of this analysis.}
\label{fig_CL_sens}
\end{figure}

 The sensitivity of the SK solar neutrino data to find a true periodicity is tested using a pseudo-experiment study. 
One thousand MC experiments are generated to simulate the five-day interval solar neutrino flux for each frequency (from $4\times10^{-4}$ days$^{-1}$ to 0.2 days$^{-1}$) and various modulation amplitudes. The power for each MC sample is calculated using a log-likelihood method. Figure\,\ref{fig_CL_sens} shows the sensitivity of finding the true period for given amplitudes and periods with 95\% confidence level (CL). The false alarm criteria $P_L$ is larger than $19.48$ for a 98\% chance of rejecting the null hypothesis. The sensitivity varies rapidly near $~0.2$ days$^{-1}$ close to the sampling frequency. For frequencies greater than $0.2$ days$^{-1}$, amplitude fluctuation is canceled when the flux is averaged over a five-day interval. 
	The results indicate that the solar neutrino flux modulation is ruled out for amplitudes greater than 5.1\% with 95\% CL (or ruled out greater than 5.2\% amplitude with 90\% CL for a 99\% chance of rejecting the null hypothesis) in the frequency range of $<0.15$ days$^{-1}$. The result is significantly improved compared to SK-I result (ruled out 10\% amplitude) \,\cite{modu_sk1}. It is a factor 2 stronger constraint than the results from SNO (10.5\% amplitude, 90\% CL and 99\% false alarm)\,\cite{PhysRevD.72.052010} as summarized in Table \ref{tab_comp_bosksn}.

\par Assuming the modulation of solar neutrino flux is only attributed to variations of the solar core temperature, as $\phi(\nu) = \alpha T^{25}$ where the $\alpha$ is a coefficient, the stability of the solar core temperature is $\delta T / T \simeq (1/25)\,\delta\phi(\nu)/\phi(\nu)$. Under this assumption, our result implies that the solar core temperature has less than 0.2\% variations, which can be compared to the SK-I result of 0.4\%.\\

\par The RSFP process and Mikheyev-Smirnov-Wolfenstein effect in a radiation zone or a tachocline of the Sun may cause potential modulations of solar neutrino flux. The probability of electron neutrino disappearance is numerically estimated with an exponent reduction factor for a magnetic field $B$ and a magnetic moment $\mu$, 
\begin{equation*}
P_{\nu_e \rightarrow \nu_e} = \frac{1}{2}\left[1 - \cos{2\theta_v}  \left( 1 - 2 P(\mu B=0)e^{-C\pi(\mu B)^{2}}\right)\right],
\end{equation*}
\noindent where $\theta_v$ is the vacuum mixing angle, $P$ is a hopping probability of mass eigenstates in the passage through the resonance region\,\cite{PhysRevD.72.033008}. 
The $C$ consists of the squared neutrino mass difference, neutrino energy, and a constant exponent proportional to the electron density \cite{PhysRevD.54.6323}. 
For the large mixing angle solution, with an average solar magnetic field of 10$^{5}$\,G and a neutrino's magnetic moment less than 10$^{-11}\mu_{\text{B}}$, flux modulation is expected to be less than 2\,\%\,\cite{PhysRevD.72.033008, CALDWELL2005543}, and the sole RSFP effect in solar amplitude is less than 0.1\,\% \cite{PhysRevD.72.033008}. These are yet to be probed in future experiments.

\par In the former study in reference\,\cite{modu_sk1}, we searched for solar neutrino flux modulations using the Lomb-Scargle periodogram method\,\cite{1976Ap_SS_Lomb,1982ApJ_Scar}. We performed the period search using the same method to compare the results directly. 

In the Lomb-Scargle method, the normalized Lomb power is given by 
\begin{equation*}
\label{eqn:lomb}
\begin{split}
P_{N}(\omega)\equiv \frac{1}{2\sigma^{2}}\biggl(& \frac{[\Sigma_{j}(\phi_{j}-\overline{\phi})\cos\,\omega(t_{j}-\tau)]^{2}}{\Sigma_{j}\cos^{2}\omega(t_{j}-\tau)} \\
& + \frac{[\Sigma_{j}(\phi_{j}-\overline{\phi})\sin\,\omega(t_{j}-\tau)]^{2}}{\Sigma_{j}\sin^{2}\omega(t_{j}-\tau)}\biggl),
\end{split}
\end{equation*}

\noindent where the measured flux in the \(j\)th bin is \(\phi_{j}\). The average time with live time weight in the \(j\)th bin is \(t_{j}\), \(\omega\) is the frequency for the test, \(\overline{\phi}\equiv\Sigma_{i=1}^{N}\phi_{i}/N\), \(\sigma^2\equiv\Sigma_{i=1}^N(\phi_{i}-\overline\phi)^{2}/(N-1)\), and offset time \(\tau\) is defined by \(\tan(2\omega \tau)\equiv (\Sigma_{j}\sin2\omega t_{j})/(\Sigma_{j}\cos2\omega t_{j})\). 
We scanned 100000 frequencies in the range of $10^{-6}$ days\(^{-1}\) to 0.2 days\(^{-1}\). A maximum Lomb power of 9.39 is observed at the frequency 0.143 days\(^{-1}\). To evaluate CL, 10000 MC experiments are generated. Random Gaussian fluctuations at a measured mean flux value and the measured standard deviations are used as the error values. The mean of gaussian is $2.36\times10^{6}$\,cm$^{-2}$s$^{-1}$ and the error is $0.486\times10^{6}$\,cm$^{-2}$s$^{-1}$. 34.3\% of these MC experiments exceeding maximum power 9.39. 
Therefore, we found no significant periodic modulation in the five-day solar neutrino sample using the Lomb-Scargle method. We tested the generalized Lomb-Scargle method to consider the errors in each data point \cite{GLS2009}, and found no modulation either.

\begin{figure}
\centering
\includegraphics[width=1\linewidth]{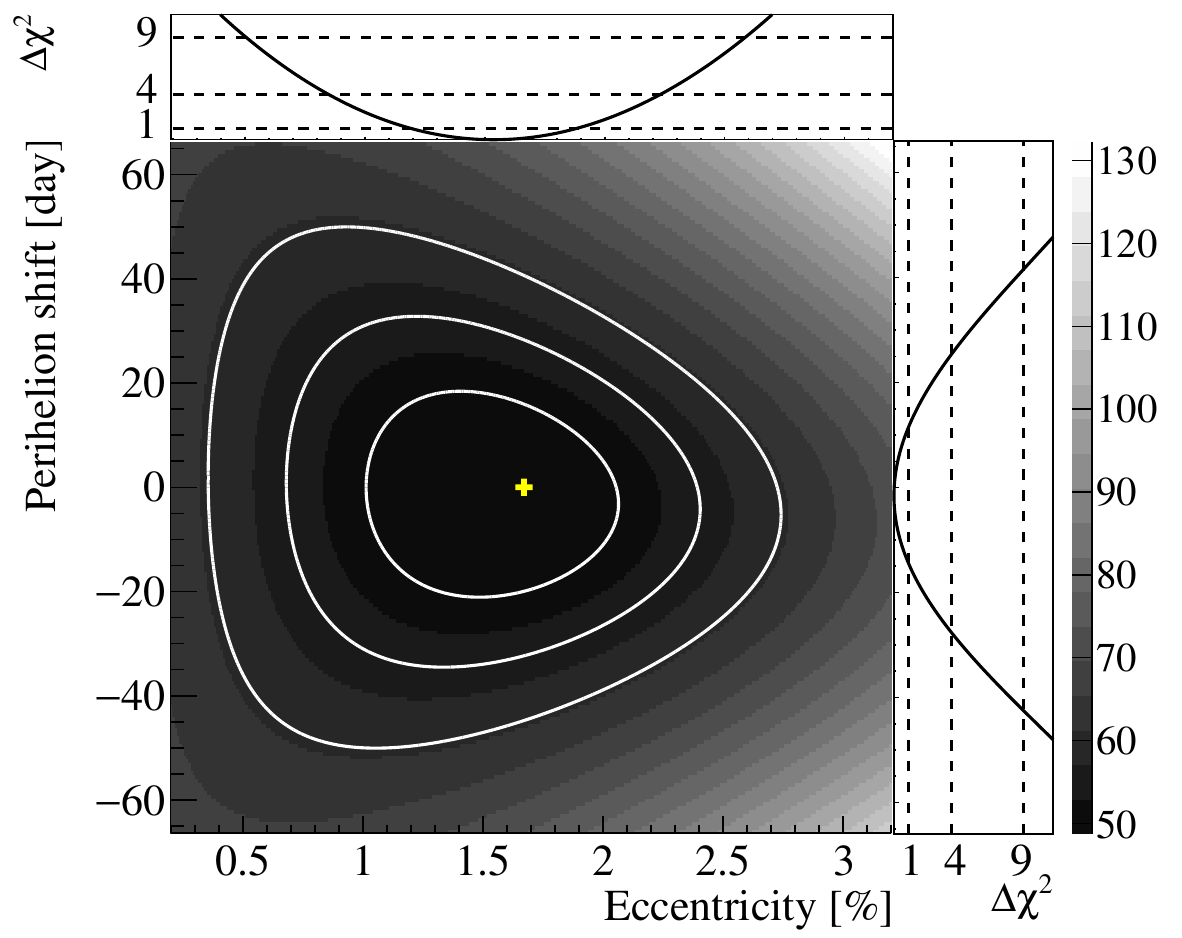}
\caption{
\(\Delta\chi^{2}\) for Kepler constants (\(\epsilon\), \(t_{p}\)) with contours \(\Delta\chi^{2}\)=$1\sigma, 2\sigma, 3\sigma$; \(\epsilon=(1.53\pm0.35)\,\%\) (expected value\,=\,1.67\%, yellow cross), \(\delta t_{\text{peri}}=(-1.5\pm13.5)\) days shifted from Jan 3rd with minimum \(\chi^{2}/NDF=48.74/(48-3)\).
}
\label{fit_sea_flux}
\end{figure}

{{\it Annual modulation}.---}We test the annual variation of the solar neutrino flux, which changes by about 7\,\% due to the Kepler orbital eccentricity of the Earth around the Sun. We divide the solar neutrino data into 12 bins for each SK run period.
Each bin width corresponds to 365.25/12, and all years for the SK period are combined.
For example, the first bin of SK-IV includes 10 months of statistics for a 10-year data taking. Figure \ref{sea_flux} shows the distribution for each SK phase. To check consistency with astronomical data, we minimize a $\chi^2$ for the solar neutrino flux at 1 A.U., the eccentricity $\epsilon$ and the perihelion shift $\delta t_{\text{peri}}$ to the astronomically observed perihelion $t_{\text{peri}}$:

\begin{figure}[t!]
\centering
\includegraphics[width=0.95\linewidth]{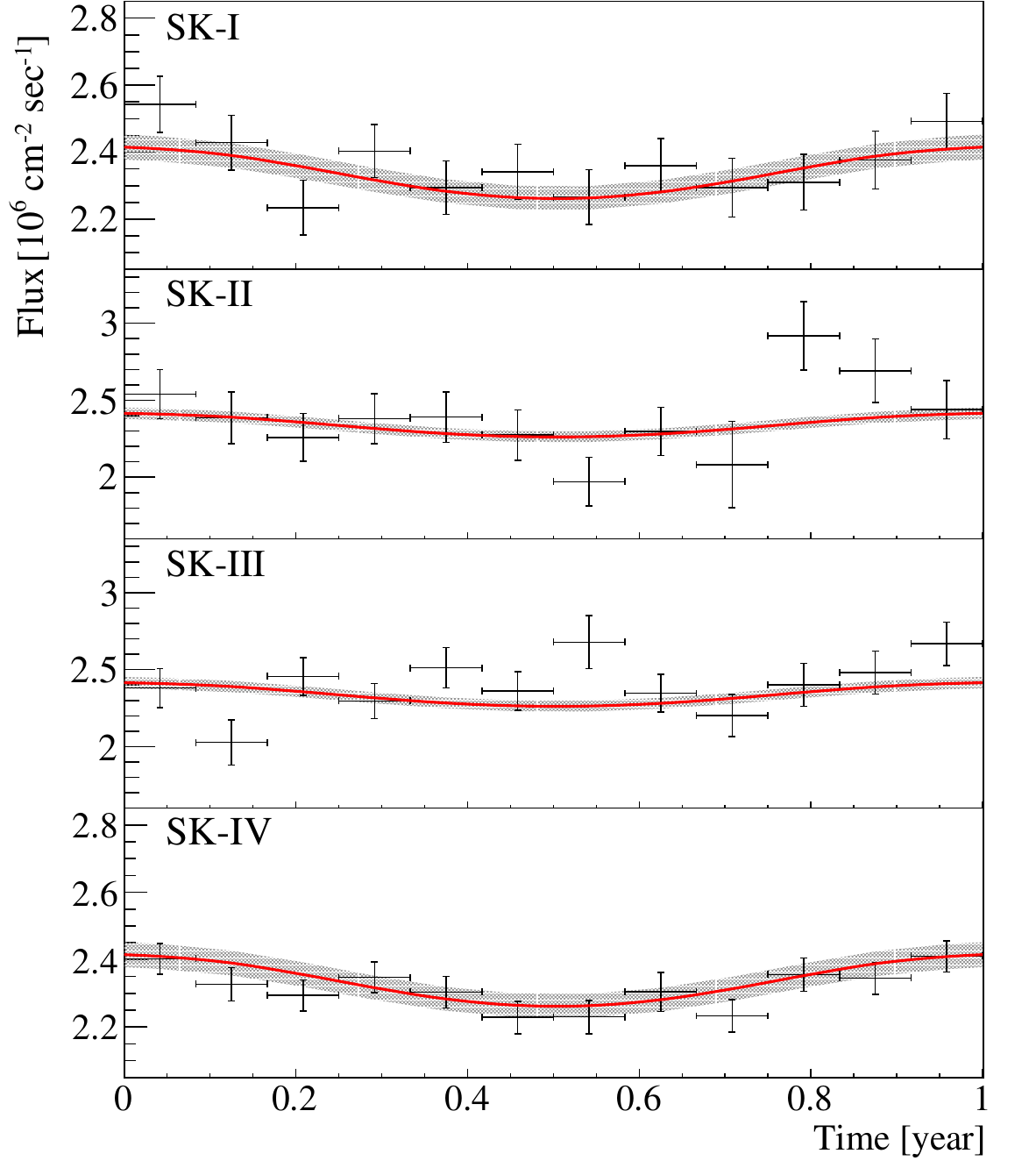}
\caption{The seasonal variation of the solar neutrino flux for SK-I/II/III/IV. The red curve and shade are flux and error in best fit, $f=(2.335\pm0.036)\times10^{6}$ \,cm$^{-2}$s$^{-1}$ at 1 A.U. times inverse of squared distance between the Sun and the Earth.}
\label{sea_flux}
\end{figure}

\begin{equation*}\label{eqn:chi_kep}
\begin{array}{ll}
\chi^{2}=\sum_{p}\left[\sum^{12}_{b=1}\left(\frac{D_{p,b}-E_{p,b}(f,\epsilon,\delta t_{\text{peri}},\delta_{p})}{\sigma^{\text{stat}}_{p,b}} \right)^{2}+\delta_{p}^{2}\right],
\\
\end{array}
\end{equation*}

\begin{equation*}\label{eqn:chi_kep_E}
E_{p,b}=\frac{f\left(1-\delta_{p}\sigma_{p}^{\text{flux\,\,syst}}\right)}{r_{b}^{2}(\epsilon,t_{\text{peri}}+\delta t_{\text{peri}})},
\end{equation*}
\noindent where $D_{p,b}$ is total solar neutrino flux for each SK run $p =$ I, II, III, IV and month $b=1\sim12$. $E_{p,b}$ is the expected flux with statistical error $\sigma^{\text{stat}}_{p,b}$. $r_{b}$ is the distance from the Sun to the Earth, and $f$ is the flux at 1 A.U. The total number of bins is 48 with three free parameters $(f, \epsilon, \delta t_{\text{peri}})$, four nuisance parameters $\delta_{p}$, and four systematic errors $\sigma_{p}^{\text{flux\,\,syst}}$. The systematic errors are different because the SK detector is renovated for each period, such as installation of more photomultiplier tubes, change of data acquisition system, etc\,\cite{sk1_full,sk2_full,sk3_full,sk4_full}. The $\chi^2$ minimization is carried out by scanning a {400 $\times$ 400} grid ($\epsilon$, $\delta t_{\text{peri}}$) parameter space. Figure \ref{fit_sea_flux} shows the $\Delta \chi^2$ contour of eccentricity and perihelion parameters. 

\par The best fit of eccentricity is $\epsilon=(1.53\pm0.35)\%$ which is consistent with the astronomical observation of $1.67\%$. The best fit of perihelion date shifted \(\delta t_{\text{peri}}=(-1.5\pm13.5)\) days from the expected date (January 3). The minimum \(\chi^{2}\)/NDF (number of d.o.f.) is \(48.74/(48-3)\). The measured average solar neutrino flux at 1 A.U. is $f=(2.335\pm0.036)\times10^{6}$\,cm$^{-2}$s$^{-1}$, where four nuisance parameters, $\delta_{\text{\,I,\,II,\,III,\,IV}}$, for the best fit are $-0.23, -0.16, -0.60, 0.62$. The observed annual modulation of the solar $^8$B neutrino flux is consistent with a flux proportional to the inverse of the squared distance between the Sun and the Earth. This is consistent with the other results of solar neutrino experiments. Borexino experiment measures $\epsilon=(1.84\pm0.32)\%,~\delta t_{\text{peri}}=(7\pm20)$ days from the seasonal flux of $^7$Be solar neutrino flux data\,\cite{APPEL2023102778}. The SNO experiment measures $\epsilon=(1.43\pm0.86)\,\%$\,\cite{PhysRevD.72.052010}. The results are summarized in Table\,\ref{tab_comp_bosksn}.  

{{\it Conclusion.---}}We presented a time variation study of solar neutrino flux at Super-Kamiokande over 5804 days (SK-I, II, III, IV). The flux of solar neutrinos exhibits annual modulation. The Kepler constants for the Earth-Sun orbit are evaluated. The measured eccentricity (1.53$\pm$0.35)\% and perihelion shift ($-$1.5\(\pm13.5\)) days are consistent with astronomical measurements.
The potential periodic modulation is tested with five-day interval solar neutrino data using the likelihood and Lomb-Scargle periodogram methods. The results indicate that there is no significant solar neutrino flux modulation greater than 5.1\% ($<0.15$ days$^{-1}$) of modulation amplitude.


\par We gratefully acknowledge the cooperation of the Kamioka Mining and Smelting Company. The Super-Kamiokande experiment has been built and operated from funding by the Japanese Ministry of Education, Culture, Sports, Science and Technology; the U.S. Department of Energy; and the U.S. National Science Foundation. Some of us have been supported by funds from the National Research Foundation of Korea (NRF-2009-0083526, NRF-2022R1A5A1030700, NRF-2022R1A3B1078756) funded by the Ministry of Science, Information and Communication Technology (ICT); the Institute for Basic Science (IBS-R016-Y2); and the Ministry of Education (2018R1D1A1B07049158, 2021R1I1A1A01042256, 2021R1I1A1A01059559); the Japan Society for the Promotion of Science; the National Natural Science Foundation of China under Grants No. 12375100; the Spanish Ministry of Science, Universities and Innovation (grant PID2021- 124050NB-C31); the Natural Sciences and Engineering Research Council (NSERC) of Canada; the Scinet and Westgrid consortia of Compute Canada; the National Science Centre (UMO2018/30/E/ST2/00441 and UMO-2022/46/E/ST2/00336) and the Ministry of Science and Higher Education (2023/WK/04), Poland; the Science and Technology Facilities Council (STFC) and Grid for Particle Physics (GridPP), UK; the European Union’s Horizon 2020 Research and Innovation Programme under the Marie Sklodowska-Curie grant agreement no. 754496; H2020-MSCA-RISE-2018 JENNIFER2 grant agreement no.822070, H2020-MSCA-RISE-2019 SK2HK grant agreement no. 872549; and European Union's Next Generation EU/PRTR grant CA3/RSUE2021-00559; the National Institute for Nuclear Physics (INFN), Italy.

\bibliography{sktimevariation}
\bibliographystyle{apsrev4-1}
\end{document}

%% file: authors-20230602.tex
\newcommand{\AFFicrr}{\affiliation{Kamioka Observatory, Institute for Cosmic Ray Research, University of Tokyo, Kamioka, Gifu 506-1205, Japan}}
\newcommand{\AFFkashiwa}{\affiliation{Research Center for Cosmic Neutrinos, Institute for Cosmic Ray Research, University of Tokyo, Kashiwa, Chiba 277-8582, Japan}}
\newcommand{\AFFipmu}{\affiliation{Kavli Institute for the Physics and
Mathematics of the Universe (WPI), The University of Tokyo Institutes for Advanced Study,
University of Tokyo, Kashiwa, Chiba 277-8583, Japan }}
\newcommand{\AFFmad}{\affiliation{Department of Theoretical Physics, University Autonoma Madrid, 28049 Madrid, Spain}}
\newcommand{\AFFubc}{\affiliation{Department of Physics and Astronomy, University of British Columbia, Vancouver, BC, V6T1Z4, Canada}}
\newcommand{\AFFbu}{\affiliation{Department of Physics, Boston University, Boston, MA 02215, USA}}
\newcommand{\AFFuci}{\affiliation{Department of Physics and Astronomy, University of California, Irvine, Irvine, CA 92697-4575, USA }}
\newcommand{\AFFcsu}{\affiliation{Department of Physics, California State University, Dominguez Hills, Carson, CA 90747, USA}}
\newcommand{\AFFcnm}{\affiliation{Institute for Universe and Elementary Particles, Chonnam National University, Gwangju 61186, Korea}}
\newcommand{\AFFduke}{\affiliation{Department of Physics, Duke University, Durham NC 27708, USA}}
\newcommand{\AFFgifu}{\affiliation{Department of Physics, Gifu University, Gifu, Gifu 501-1193, Japan}}
\newcommand{\AFFgist}{\affiliation{GIST College, Gwangju Institute of Science and Technology, Gwangju 500-712, Korea}}
\newcommand{\AFFuh}{\affiliation{Department of Physics and Astronomy, University of Hawaii, Honolulu, HI 96822, USA}}
\newcommand{\AFFicl}{\affiliation{Department of Physics, Imperial College London , London, SW7 2AZ, United Kingdom }}
\newcommand{\AFFkek}{\affiliation{High Energy Accelerator Research Organization (KEK), Tsukuba, Ibaraki 305-0801, Japan }}
\newcommand{\AFFkobe}{\affiliation{Department of Physics, Kobe University, Kobe, Hyogo 657-8501, Japan}}
\newcommand{\AFFkyoto}{\affiliation{Department of Physics, Kyoto University, Kyoto, Kyoto 606-8502, Japan}}
\newcommand{\AFFliv}{\affiliation{Department of Physics, University of Liverpool, Liverpool, L69 7ZE, United Kingdom}}
\newcommand{\AFFmiyagi}{\affiliation{Department of Physics, Miyagi University of Education, Sendai, Miyagi 980-0845, Japan}}
\newcommand{\AFFnagoya}{\affiliation{Institute for Space-Earth Environmental Research, Nagoya University, Nagoya, Aichi 464-8602, Japan}}
\newcommand{\AFFkmi}{\affiliation{Kobayashi-Maskawa Institute for the Origin of Particles and the Universe, Nagoya University, Nagoya, Aichi 464-8602, Japan}}
\newcommand{\AFFpol}{\affiliation{National Centre For Nuclear Research, 02-093 Warsaw, Poland}}
\newcommand{\AFFsuny}{\affiliation{Department of Physics and Astronomy, State University of New York at Stony Brook, NY 11794-3800, USA}}
\newcommand{\AFFokayama}{\affiliation{Department of Physics, Okayama University, Okayama, Okayama 700-8530, Japan }}
\newcommand{\AFFosaka}{\affiliation{Department of Physics, Osaka University, Toyonaka, Osaka 560-0043, Japan}}
\newcommand{\AFFox}{\affiliation{Department of Physics, Oxford University, Oxford, OX1 3PU, United Kingdom}}
\newcommand{\AFFqmul}{\affiliation{School of Physics and Astronomy, Queen Mary University of London, London, E1 4NS, United Kingdom}}
\newcommand{\AFFregina}{\affiliation{Department of Physics, University of Regina, 3737 Wascana Parkway, Regina, SK, S4SOA2, Canada}}
\newcommand{\AFFseoul}{\affiliation{Department of Physics, Seoul National University, Seoul 151-742, Korea}}
\newcommand{\AFFsheff}{\affiliation{Department of Physics and Astronomy, University of Sheffield, S3 7RH, Sheffield, United Kingdom}}
\newcommand{\AFFshizuokasc}{\affiliation{Department of Informatics in
Social Welfare, Shizuoka University of Welfare, Yaizu, Shizuoka, 425-8611, Japan}}
\newcommand{\AFFstfc}{\affiliation{STFC, Rutherford Appleton Laboratory, Harwell Oxford, and Daresbury Laboratory, Warrington, OX11 0QX, United Kingdom}}
\newcommand{\AFFskk}{\affiliation{Department of Physics, Sungkyunkwan University, Suwon 440-746, Korea}}
\newcommand{\AFFtodai}{\affiliation{Department of Physics, University of Tokyo, Bunkyo, Tokyo 113-0033, Japan }}
\newcommand{\AFFtit}{\affiliation{Department of Physics,Tokyo Institute of Technology, Meguro, Tokyo 152-8551, Japan }}
\newcommand{\AFFtus}{\affiliation{Department of Physics, Faculty of Science and Technology, Tokyo University of Science, Noda, Chiba 278-8510, Japan }}
\newcommand{\AFFtoronto}{\affiliation{Department of Physics, University of Toronto, ON, M5S 1A7, Canada }}
\newcommand{\AFFtriumf}{\affiliation{TRIUMF, 4004 Wesbrook Mall, Vancouver, BC, V6T2A3, Canada }}
\newcommand{\AFFtokai}{\affiliation{Department of Physics, Tokai University, Hiratsuka, Kanagawa 259-1292, Japan}}
\newcommand{\AFFtsinghua}{\affiliation{Department of Engineering Physics, Tsinghua University, Beijing, 100084, China}}
\newcommand{\AFFynu}{\affiliation{Department of Physics, Yokohama National University, Yokohama, Kanagawa, 240-8501, Japan}}
\newcommand{\AFFllr}{\affiliation{Ecole Polytechnique, IN2P3-CNRS, Laboratoire Leprince-Ringuet, F-91120 Palaiseau, France }}
\newcommand{\AFFbari}{\affiliation{ Dipartimento Interuniversitario di Fisica, INFN Sezione di Bari and Universit\`a e Politecnico di Bari, I-70125, Bari, Italy}}
\newcommand{\AFFnapoli}{\affiliation{Dipartimento di Fisica, INFN Sezione di Napoli and Universit\`a di Napoli, I-80126, Napoli, Italy}}
\newcommand{\AFFroma}{\affiliation{INFN Sezione di Roma and Universit\`a di Roma ``La Sapienza'', I-00185, Roma, Italy}}
\newcommand{\AFFpadova}{\affiliation{Dipartimento di Fisica, INFN Sezione di Padova and Universit\`a di Padova, I-35131, Padova, Italy}}
\newcommand{\AFFkeio}{\affiliation{Department of Physics, Keio University, Yokohama, Kanagawa, 223-8522, Japan}}
\newcommand{\AFFwinnipeg}{\affiliation{Department of Physics, University of Winnipeg, MB R3J 3L8, Canada }}
\newcommand{\AFFkcl}{\affiliation{Department of Physics, King's College London, London, WC2R 2LS, UK }}
\newcommand{\AFFwarwick}{\affiliation{Department of Physics, University of Warwick, Coventry, CV4 7AL, UK }}
\newcommand{\AFFral}{\affiliation{Rutherford Appleton Laboratory, Harwell, Oxford, OX11 0QX, UK }}
\newcommand{\AFFwu}{\affiliation{Faculty of Physics, University of Warsaw, Warsaw, 02-093, Poland }}
\newcommand{\AFFbcit}{\affiliation{Department of Physics, British Columbia Institute of Technology, Burnaby, BC, V5G 3H2, Canada }}
\newcommand{\AFFtohoku}{\affiliation{Department of Physics, Faculty of Science, Tohoku University, Sendai, Miyagi, 980-8578, Japan }}
\newcommand{\AFFicise}{\affiliation{Institute For Interdisciplinary Research in Science and Education, ICISE, Quy Nhon, 55121, Vietnam }}
\newcommand{\AFFilance}{\affiliation{ILANCE, CNRS - University of Tokyo International Research Laboratory, Kashiwa, Chiba 277-8582, Japan}}
\newcommand{\AFFibs}{\affiliation{Center for Underground Physics, Institute for Basic Science (IBS), Daejeon, 34126, Korea}}
\newcommand{\AFFglasgow}{\affiliation{School of Physics and Astronomy, University of Glasgow, Glasgow, Scotland, G12 8QQ, United Kingdom}}
\newcommand{\AFFoecu}{\affiliation{Media Communication Center, Osaka Electro-Communication University, Neyagawa, Osaka, 572-8530, Japan}}

\AFFicrr
\AFFkashiwa
\AFFmad
\AFFbu
\AFFbcit
\AFFuci
\AFFcsu
\AFFcnm
\AFFduke
\AFFllr
\AFFgifu
\AFFgist
\AFFglasgow
\AFFuh
\AFFibs
\AFFicise
\AFFicl
\AFFbari
\AFFnapoli
\AFFpadova
\AFFroma
\AFFilance
\AFFkeio
\AFFkek
\AFFkcl
\AFFkobe
\AFFkyoto
\AFFliv
\AFFmiyagi
\AFFnagoya
\AFFkmi
\AFFpol
\AFFsuny
\AFFokayama
\AFFoecu
\AFFox
\AFFral
\AFFseoul
\AFFsheff
\AFFshizuokasc
\AFFstfc
\AFFskk
\AFFtohoku
\AFFtokai
\AFFtodai
\AFFipmu
\AFFtit
\AFFtus
\AFFtriumf
\AFFtsinghua
\AFFwu
\AFFwarwick
\AFFwinnipeg
\AFFynu

\author{K.~Abe}
\AFFicrr
\AFFipmu
\author{C.~Bronner}
\AFFicrr
\author{Y.~Hayato}
\AFFicrr
\AFFipmu
\author{K.~Hiraide}
\AFFicrr
\AFFipmu
\author{K.~Hosokawa}
\AFFicrr
\author{K.~Ieki}
\author{M.~Ikeda}
\AFFicrr
\AFFipmu
\author{J.~Kameda}
\AFFicrr
\AFFipmu
\author{Y.~Kanemura}
\author{R.~Kaneshima}
\author{Y.~Kashiwagi}
\AFFicrr
% ...


\collaboration{The Super-Kamiokande Collaboration}
\noaffiliation